\input epsf.sty
\documentclass{aa}
\usepackage{deluxetable}
\begin{document}
\def\RS{R_{\rm Schw}}
\def\Agata{R\' o\. za\' nska~}

\title{Constraints for the accretion disk evaporation rate in AGN 
from the existence of the Broad Line Region}

\titlerunning{Disk evaporation and the BLR}

\author{ B. Czerny, \inst{1} 
           A. R\' o\. za\' nska,  
          \inst{1}
           J. Kuraszkiewicz
          \inst{2}
}

\institute{$^1$Copernicus Astronomical Center, Bartycka 18, 00-716 
    Warsaw, Poland \\
   $^2$ Harvard-Smithsonian Center for Astrophysics, 60 Garden Street, MS83,
   Cambridge,MA, USA \\
}

 \date{Received ...; accepted ...}

\abstract{We analyze the consequences of the hypothesis that the formation of 
the Broad Line Region is intrinsically connected to the existence of the
cold accretion disk. We assume that the Broad Line Region radius is reliably 
estimated by the formula of Kaspi et al. (2000). We consider three models
of the disappearance of the inner disk that limit the existence of the 
Broad Line Region: (i) the classical ADAF approach, i.e. the inner hot flow 
develops whenever it can exist (ii) the disk evaporation model of Meyer \& 
Meyer-Hofmeister (2002) (iii) the generalized disk evaporation model of 
R\' o\. za\' nska \& Czerny (2000b). For each of the models, we determine the
minimum value of the Eddington ratio and the maximum value of the broad line 
widths as functions of the viscosity parameter $\alpha$ and the magnetic
field parameter $\beta$. We compare the predicted 
parameter space with observations of several AGN. 
Weak dependence of the maximum value of the FWHM and minimum 
value of the Eddington ratio on the black hole mass in our sample is 
noticeable. It seems to favor the description of the cold disk/hot inner 
flow transition 
as in the classical ADAF approach rather than with the model
of disk evaporation due to conduction between the disk and accreting corona.

}

\maketitle

\section{Introduction}

One of the most important issues in studying accretion onto 
compact objects is to understand the process that causes some fraction of the 
accreting material to be 
heated sufficiently strongly 
to form a hot optically thin, predominantly thermal 
plasma in the vicinity of a black hole (see e.g. Narayan, Mahadevan \& 
Quataert 1998;  Collin et al. 2001;  Czerny 2003).  

Observations of the X-ray emission from this plasma do not uniquely
constrain the geometry of the plasma. Several phenomenological models are 
under consideration;  those include 
magnetic flares above an accretion disk extending
always down to the marginally stable orbit (Galeev, Rosner \& Vaiana 1979;
Haardt, Maraschi \& Ghisellini 1994), or 
the solutions of accretion disk structure with inner optically 
thin (possible advection-dominated) flow (Ichimaru 1977; Narayan \& Yi 1994,
Abramowicz et al. 1995). The geometry possibly depends on the luminosity 
to the Eddington luminosity ratio in a given object.  
Detailed XMM-Newton and Chandra studies of the X-ray 
reflection (both continuum and iron $K_{\alpha}$ line) suggest that at high 
Eddington ratios the disk extends to the 
marginally stable orbit and its surface appears 
strongly ionized (e.g. \Agata et al.
2004 for Ton S180; Janiuk et al. 2001 and  Pounds et al. 2003 for PG1211+143;
Sako et al. 2003 for MCG -6-30-15 and Markarian 766) 
while in low Eddington ratio sources the spectra are well-described as 
being due to cold 
reflection from the outer part of the accretion disk (e.g. O'Brien 
et al. 2003). 

Those observations motivate the development of models in which the 
material slowly evaporates from the cold disk into a hot corona as 
the black hole is approached (Liu et al. 1999, 2002, \Agata \& Czerny 2000b, 
Meyer \& Meyer-Hofmeister 2002, Meyer-Hofmeister \& Meyer 2003). 
All such models predict that at high Eddington ratios
the evaporation effect is relatively inefficient so the 
cold optically thick disk survives all the way down to the black hole while for
low Eddington ratio 
sources the evaporation leads to a complete evacuation of the 
disk and formation of an inner optically thin, hot flow.

However, the physical description of the evaporation rate involves a number 
of complicated processes, and various models 
rely on several underlying assumptions so it is
very important to verify the models against observational constraints.
In this paper we use the constraint based on the measured width of the broad 
emission lines. The exact nature of the broad line region (BLR) is not known 
but most models connect the formation of low ionization lines like $H_{\beta}$
with the wind developing close to the disk surface (Collin-Souffrin et al. 
1988; Dultzin-Hacyan et al. 2000, Laor 2003a). If such an approach is
adopted, the existence of the broad lines at a certain range of radii is 
possible only if the cold disk extends also at least to the same radial range,
and regions of the efficient line/wind formation are related to specific
phenomena taking place within the disk. For example, such an approach led 
Elvis (2000) to propose a general, consistent scheme for quasar structure.

The details of the processes, including disk/BLR coupling, are by no means 
clear but the large amount of the existing data opens possibilities 
to test various hypotheses of the nature of this relation. Laor (2003b) 
noticed that there appears to be an upper limit to the broad line width
of about 25 000 km/s which apparently sets a lower limit for the luminosity
of the broad line object. He postulated that a significant population
of AGN exists that intrinsically do not have broad lines.
Further developing this idea, Nicastro et al.
(2003) found the division between the AGN with and without broad lines 
occurs at the Eddington ratio of $\sim 0.001$, and interpreted this result
as an argument for a connection between the onset of a wind 
from the disk surface 
and the transition from a gas dominated to a radiation pressure dominated
part of the cold disk.   

In this paper we assume that the outer cold disk transitions to 
an inner hot, optically thin and radiatively inefficient flow.  We further
assume that the BLR may form above a cold disk, in its atmosphere or
outflowing wind, but not above a hot inner flow. These assumptions allow
us to qualitatively test the theoretical predictions for the location of 
such a transition. We consider three specific models: the classical 
condition for the existence of ADAF 
(Abramowicz et al. 1995), the evaporation 
model of Meyer \& Meyer-Hofmeister (2002) and the evaporation model of \Agata
and Czerny (2000) generalized for the case of a non-negligible magnetic field. 

\section{Method}

\subsection{Size of the BLR}

The size of the broad line region depends predominantly on the luminosity of
the active nucleus. The most accurate results are based on the reverberation 
method. Delays of the $H_{\beta}$ response with respect to the changes of the
continuum lead to the relation (Kaspi et al. 2000) 
\begin{equation}
  R_{BLR} = 32.9 \left({\lambda L_\lambda (5100 \AA) \over 10^{44}}
\right)^{0.7}~ {\rm lt~days},
\label{eq:Kaspi}
\end{equation}
with a relatively high correlation coefficient in the luminosity 
range covering over three decades (with most sources between $10^{42}$ and
$10^{45}$ erg s$^{-1}$). 

The value of the monochromatic luminosity at 5100~{\AA } 
is surprisingly closely related to the bolometric luminosity of an AGN 
\begin{equation}
  L_{bol} = 9 \lambda L_{\lambda}(5100 \AA).
\label{eq:bolo}
\end{equation}
(e.g. Kaspi et al. 2000, Collin et al. 2002). 

The bolometric luminosity can
be measured in dimensionless Eddington units, where
\begin{equation}
L_{Edd} = {4 \pi GMm_p c \over \sigma_T}.
\label{eq:Edd}
\end{equation}

We express the mass 
in units of $10^8 M_{\odot}$ (i.e. $M_8 = M/10^8 M_{\odot}$). 
Finally, we give the BLR radius in the units of Schwarzschild radius,
$R_{Schw} = 2 GM/c^2$, since theoretical estimates usually refer to 
such a quantity:
\begin{equation}
R_{BLR} = 1.8 \times 10^4  \left( \frac {L_{bol}}{L_{Edd}} \right)^{0.7} 
 M_8^{-0.3} ~~R_{Schw}.
\label{eq:BLR}
\end{equation}

\subsection{Line width}

The determination of the black hole mass by Kaspi et al. (2000) was based on
the assumption that the motion of the BLR clouds is predominantly Keplerian.
This approach was supported by a separate analysis of the several emission 
lines in NGC 5548 by Peterson \& Wandel (1999).  To determine the FWHM 
from the Keplerian velocity the correction factor $2/\sqrt 3$
is needed, which accounts for velocities in three dimensions and for the full
width being two times higher than the velocity dispersion, i.e.: 
\begin{equation}
  FWHM = {2 \over \sqrt{3}} \sqrt{GM \over R_{BLR}}.
\label{eq:FWHM1}
\end{equation}

Here, we reverse the approach of Kaspi at al. (2000), and we express the 
width of the $H_{\beta}$ line as a function of the black hole mass 
and its luminosity
 \begin{equation}
   FWHM = 1824 ~ M_8^{0.15} ~\left( \frac {L_{bol}}{L_{Edd}} \right)^{-0.35}~~
[{\rm km ~~ s}^{-1}].
\label{eq:FWHM}
\end{equation}

\subsection{Evaporation radius}
\label{sect:evap}

The location of the transition from the optically thick disk embedded in a hot 
corona into a single phase inner ADAF depends on the assumptions underlying
the evaporation process. The transition radius, in Schwarzschild units, 
depends on the dimensionless accretion rate defined as
\begin{equation}
\dot m = {\dot M \over \dot M_{Edd}}; ~~~~ \dot M_{Edd} = {4 \pi GMm_p  \over \sigma_T c \eta_0},
\label{eq:mdot}
\end{equation}
where $\eta_0 = 1/12$ is the efficiency of disk accretion in the Newtonian
approximation.

In classical papers about the advection dominated flows (ADAFs), where
the process of disk evaporation was not considered, the 
estimate of the truncation radius
was simply based on the absence of a strong ADAF solution above this radius
(Abramowicz et al. 1995, Honma 1996, Kato \& Nakamura 1998).: 
\begin{equation}
  R_{evap}^{A} = 2.0 \dot m^{-2} \alpha_{0.1}^4 \RS.
\label{eq:evapA}
\end{equation}

The unique relation between truncation radius and mass accretion rate 
can be determined when evaporation of the cold disc is taken into account. 
In the model of a single-temperature accreting corona with the evaporation 
efficiency determined by
the conduction between the cold disk and a hot corona described as 
is customarily done for the
solar corona, this transition takes place at the evaporation radius:
\begin{equation}
  R_{evap}^* = 18.3 \dot m^{-0.85} \RS
\end{equation}
(Liu et al. 1999). This relation is derived for a single value of the 
viscosity parameter $\alpha=0.3$. 

The same model, but generalized for the presence of magnetic field within 
the accreting corona can be roughly parameterized as
\begin{equation}
  R_{evap}^{B} = 18.3 \dot m^{-0.85} \beta^{2.5}  \RS,
\label{eq:evapB}
\end{equation}
which is based on the numerical results presented  by Meyer \& Meyer-Hofmeister
(2002). Here $\beta$ is the ratio of the gas pressure to 
the total (gas + magnetic)
 pressure in the corona. Therefore, the ratio of 
gas to the magnetic pressure used in Meyer \& Meyer-Hofmeister (2002)
is $P_g/P_m=\beta/(1-\beta)$.

In the two-temperature model of accreting corona with the evaporation
efficiency calculated under a constant pressure assumption with conduction 
self-consistently determined from radiative processes 
(\Agata \& Czerny 2000a), 
this transition is predicted to be at
\begin{equation}
R_{evap}^{**}  =  19.5 \dot m^{-0.53} \alpha_{0.1}^{0.8} \RS. 
\label{eq:tran}
\end{equation}
This relation shows a different dependence on the viscosity parameter 
than that predicted in \Agata \& Czerny (2000b),
due to an error in the original paper. 

We also generalize this model for the case of the contribution of the
magnetic pressure to the
total pressure (see Appendix A).
We repeat the analysis of \Agata \& Czerny (2000b). Numerical
results for the case of low accretion rate (i.e. negligible condensation) can
be approximated analytically as 
\begin{equation}
\dot m_z = \dot m_{z}^0 \beta
\end{equation}
and consequently the evaporation radius changes to
\begin{equation}
R_{evap}^{C} =  19.5 \dot m^{-0.53} \alpha_{0.1}^{0.8} \beta^{-1.08} \RS. 
\label{eq:evapC} 
\end{equation} 

\subsection{Accretion efficiency}

\begin{figure}
\epsfxsize = 80 mm \epsfbox[20 260 530 700]{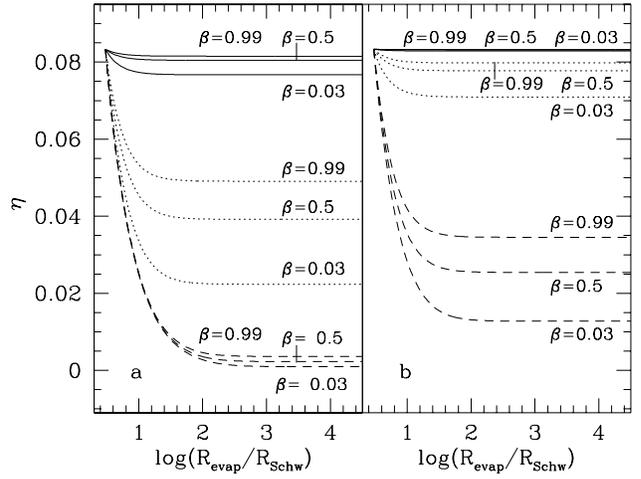}
\caption{The dependence of accretion efficiency on evaporation radius 
according to Eq.~\ref{eq:factor} for two different viscosity parameters
 $\alpha =0.1 $ (panel a), and $\alpha=0.04$ (panel b). 
 Solid lines represent the case of $\dot m= 0.3 $, dotted lines - $\dot m= 0.03$,
and the dashed lines - $\dot m= 0.003$.   }
\label{fig:ef}
\end{figure}

The bolometric luminosity is in turn related to the accretion rate through the
accretion efficiency
\begin{equation}
  L_{bol} = \eta \dot M c^2.
\end{equation}

In Eq.~\ref{eq:mdot} we used the efficiency $\eta_0$ which is appropriate 
for the description of outer 
parts of the flow well represented by a classical disk accretion.  However,
the actual efficiency of the conversion of the mass accretion 
into observed luminosity
in the innermost part of the flow may be much lower due to advection. 
Therefore, we consider the efficiency $\eta$ in a general form
\begin{equation}
  \eta = \eta_0 \cal F,
\label{eq:efifi}
\end{equation}
and we later specify $\cal F$ as a function of model parameters.
In such a general case the bolometric luminosity depends on the dimensionless 
accretion rate as:
\begin{equation}
\frac{L_{bol}}{L_{Edd}}=\dot m {\cal F}.  
\end{equation}

The existence of the evaporation radius affects the efficiency of the 
accretion. Without an inner advection-dominated flow the correction factor to 
the efficiency $\cal F$
in Eqs.~\ref{eq:BLR} and ~\ref{eq:FWHM} is equal
to 1, and the evaporation radius is equal to 3, in Schwarzschild units, 
for a non-rotating black hole. 
This efficiency drops with an increasing $R_{evap}$ since the number of soft
photons penetrating the hot region decreases with an increase of its 
radial optical depth.
For large values of $R_{evap}$ this efficiency finally saturates 
since there, the locally
emitted synchrotron and bremsstrahlung photons are the only source of soft
photons for Comptonization. According to the Monte Carlo computations 
of Kurpiewski \& Jaroszy\' nski (2000) this saturation value is on the order 
of $\varepsilon$, where:
\begin{equation}
\varepsilon=170.99 \left({\dot m \over \alpha_{0.1}^2}
 \right)^{1.5} (1.25 \beta +0.375).
\label{eq:kurpie} 
\end{equation} 

In order to approximate this complex behaviour we use a simple formula
for the correction factor mentioned above,
having in mind that $r_{evap}=R_{evap}/\RS$
\begin{equation}
  {\cal F} = {3 \over r_{evap}} + \left(1 - { 3 \over r_{evap}}\right)
{\varepsilon~ r_{evap} \over 3 + \varepsilon r_{evap}} 
  \left( {\varepsilon \over \varepsilon+1} \right),
\label{eq:factor}
\end{equation}
which gives the correct asymptotic dependence for both small and 
large evaporation radii.

The dependence of the total accretion efficiency described 
by Eqs.~\ref{eq:efifi} and ~\ref{eq:factor} on the evaporation 
radius is shown in Fig.~\ref{fig:ef}.
As requested, this efficiency reduces to 1/12 if no evaporation takes place
($r_{evap}=3$), it drops for increasing $r_{evap}$ due to the decreasing
contribution of the cold disk emission and finally it saturates at the 
value characteristic of the ADAF solution itself. This last value 
depends significantly  
both on the viscosity parameter (directly responsible for angular 
momentum transfer) and on the magnetic field strength 
parameter $\beta$ (regulating the electron cooling 
through the synchrotron emission).

\begin{figure}
\epsfxsize = 75 mm \epsfbox[50 180 560 660]{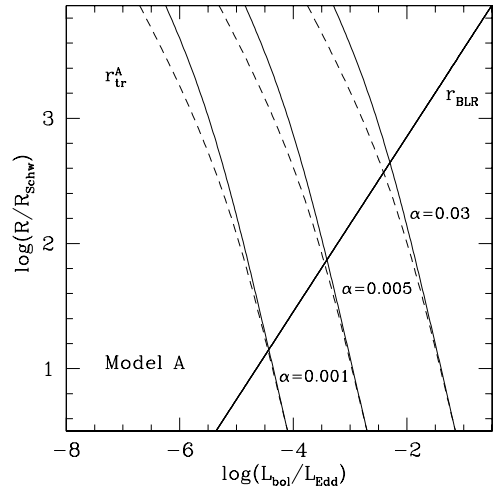}
\epsfxsize = 75 mm \epsfbox[50 180 560 660]{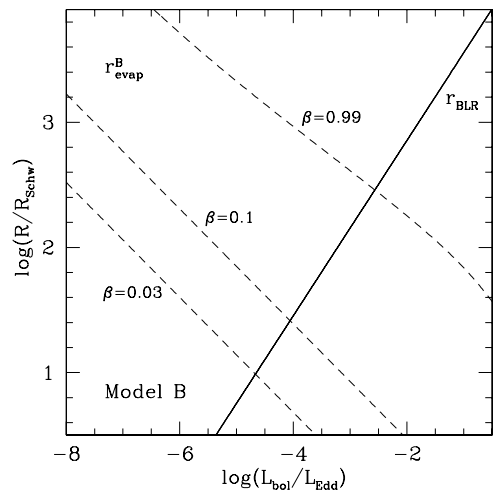}
\epsfxsize = 75 mm \epsfbox[50 180 560 660]{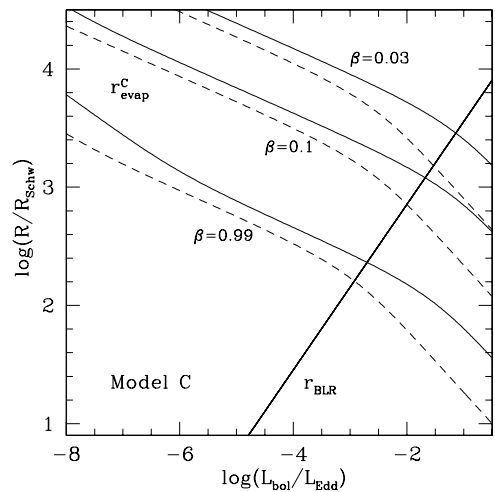}
\caption{The dependence of the BLR radius (thick continuous line) 
and the transition radius (thin lines) on the 
external accretion rate in model A (upper panel), model B (middle panel) and model C
(lower panel) for the black hole mass $10^8 M_{\odot}$.
Model A parameters: $\beta = 0.99$ (continuous thin lines),  
and $\beta=0.03$ (dashed thin lines). 
Model B parameters: 
$\alpha = 0.3$ (dashed lines). Model C parameters: 
$\alpha = 0.1$ (continuous thin lines),  
and $\alpha =0.02$ (dashed thin lines).
The intersection determines the minimum Eddington ratio 
for a broad line object
and the minimum BLR radius.}
\label{fig:radii}
\end{figure}


\subsection{Minimum Eddington ratio and maximum line width for 
broad line objects}

We assume that broad emission lines cannot form in the radial 
range inside the evaporation radius. Therefore the equality
\begin{equation}
 R_{evap} = R_{BLR} 
\end{equation}
gives the minimum Eddington ratio for which the BLR can exist 
for a given black hole mass $M$, assuming a certain value of
the viscosity parameter $\alpha$ and/or of the magnetic pressure contribution
$\beta$. 

The same condition, combined with Eq.~\ref{eq:FWHM} gives the
maximum value of the FWHM of $H_{\beta}$, again for a given $M$, $\alpha$
and/or $\beta$. 

Both FWHM of $H_{\beta}$ and the Eddington ratio can 
be obtained or estimated on the 
basis of the observational data. When 
complemented with black hole mass estimates
for each source, the measured values provide constraints for $\alpha$
and/or $\beta$. 

\section{Results}

\subsection{BLR radii and disk evaporation radii in three models}

The plots of $R_{BLR}$ and $R_{evap}$ as functions of the Eddington ratio for
the considered models are shown in Fig.~\ref{fig:radii}. The position 
of the straight line
representing $R_{BLR}$ is the same for all models and depends only on the 
mass of the object (see Eq.~\ref{eq:BLR}). The evaporation radius depends
on the model.


In the upper panel of Fig.~\ref{fig:radii} we show the results
based on the {\it strong ADAF principle} (model A) for three 
values of the viscosity 
parameter $\alpha$ and two extreme values of the parameter $\beta$. 
Lowering the value of the viscosity strongly decreases the
inner ADAF region (see Eq.~\ref{eq:evapA}). The weak dependence on $\beta$
is caused by the change in accretion flow efficiency
in the ADAF part (see Eq.~\ref{eq:kurpie}). 

The intersection of $R_{tr}^A$ and $R_{BLR}$ determines the minimum 
Eddington ratio, for a given $\alpha$ and $\beta$, for which the BLR exists: 
below this value the BLR would form in the ADAF zone.
Therefore, again for a given $\alpha$, the size of the BLR
as a function of accretion rate is defined as a fragment of the
$R_{BLR}$ line above the crossing point. For a $10^8 M_{\odot}$ object
and  $\alpha \sim 0.03$, this minimum Eddington ratio of a broad line
object is about 0.005, and the minimum value of $R_{BLR}$ is
about 400 $R_{Schw}$, corresponding to the maximum value of the FWHM equal
to 12 200 km/s. The minimum Eddington ratio increases and the FWHM maximum
decreases  with increasing
$\alpha$.

In the middle panel of Fig.~\ref{fig:radii} we plot the results for Model
B, varying the parameter $\beta$. In this case the effect of $\beta$ is
much stronger than in the ADAF approach since the magnetic field affects 
the evaporation efficiency directly (see Eq.~\ref{eq:evapB}), apart from 
the change of hot flow efficiency. A weaker magnetic field (higher $\beta$)
allows the disk to extend closer in. The minimum Eddington rate
of a broad line object is predicted to be $\sim 0.003$ for low magnetic 
field and $\sim 10^{-4}$ for $\beta = 0.1$. The lack of dependence on
$\alpha$ is caused by the arbitrary assumption of $\alpha = 0.3$ made by
Meyer \& Meyer-Hofmeister (2002), with no results available for other
values of the viscosity parameter.

The results for the model C are shown in the lower panel of 
Fig.~\ref{fig:radii}. Here we have a strong dependence on both $\alpha$
and $\beta$ since again the magnetic field plays an active role in the
process of the disk evaporation (see Eq.~\ref{eq:evapC} and the Appendix).
Generally, larger magnetic field leads to more efficient disk evaporation and 
consequently the formation of the BLR closer to the black hole is strongly
limited.

\begin{figure*}
\epsfxsize = 150 mm \epsfbox[80 520 515 690]{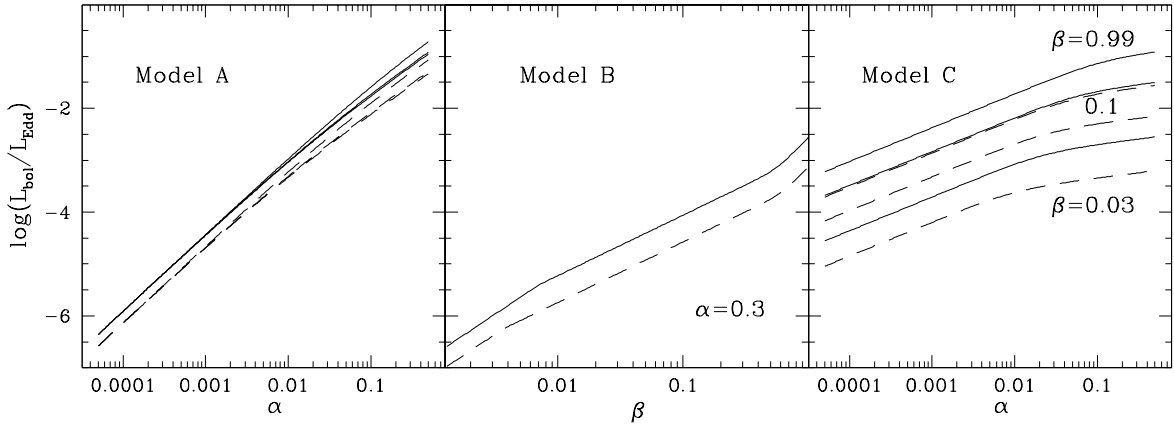}
\caption{The dependence of the minimum of the Eddington rate for three models 
A (left), B (middle), and C (right panel) on 
viscosity and magnetic parameters. 
Note that model B is computed only for $\alpha =0.3$ (see text). 
Solid lines represent the case of $M_{BH}=10^8 M_{\odot}$, while dashed 
lines -- $10^6 M_{\odot}$.}
\label{fig:Edd_limits}
\end{figure*}

\begin{figure*}
\epsfxsize = 150 mm \epsfbox[80 520 515 690]{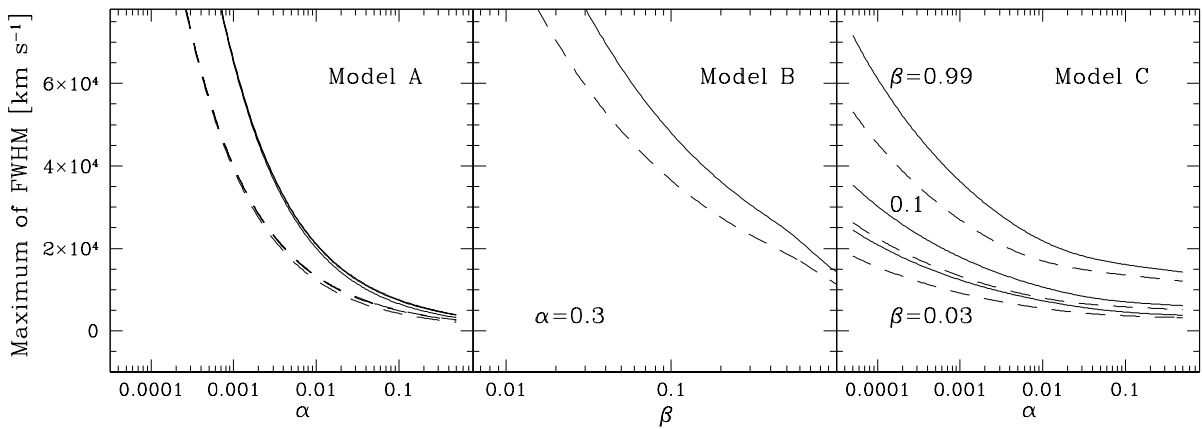}
\caption{The dependence of the maximum of FWHM for three models 
A (left), B (middle), and C (right panel) on 
viscosity and magnetic parameters. 
Note that model B is computed only for $\alpha =0.3$ (see text). 
Solid lines represent the case of $M_{BH}=10^8 M_{\odot}$, while dashed 
lines -- $10^6 M_{\odot}$.}
\label{fig:FWHM_limits}
\end{figure*}

\subsection{Minimum Eddington rate of a broad line object}

The minimum Eddington rate of a broad line object is defined
by the line crossings in Fig.~\ref{fig:radii}. It depends on the 
black hole mass, viscosity and the magnetic field. We summarize 
some of these relations in Fig~\ref{fig:Edd_limits}.

The values obtained from the classical ADAF solution are very
low unless the viscosity parameter is of the order of 0.01 or larger,
and the dependence on the magnetic field is negligible (see 
Fig~\ref{fig:Edd_limits}, left panel).  Derived 
limits formally depend on the assumed black hole mass 
(Fig.~\ref{fig:radii} was constructed for $M_{BH}=10^8 M_{\odot}$, but
the analysis can be repeated for other black hole masses).
However, in the case of the ADAF model, this dependence is rather weak.

The values obtained from model B also span the whole range, 
increasing significantly with an increase of the magnetic field. 
Larger black hole mass results in a larger value of the minimum bolometric
luminosity of a BLR object.

In the case of model C the dependence on all parameters, $\alpha$,
$\beta$ and the black hole mass is important. The Eddington ratio plot
is generally flatter, and even for very low viscosity and a strong magnetic
field the minimum Eddington ratio is larger than $\sim 10^{-5}$. 

\subsection{Maximum line width of a broad line object} 

The size of the BLR corresponds to the FWHM of the $H_{\beta}$ line 
(see Eq.~\ref{eq:FWHM1}), so we can replace the radius with the values
of FWHM which is directly measured from observations. Therefore,
the allowed region in the $R - L/L_{Edd}$ diagram translates into
the equivalent region in the ${\rm FWHM} - L/L_{Edd}$ diagram
according to Eq.\ref{eq:FWHM}. Alternatively, to represent these
limits we can plot the maximum ${\rm FWHM}$ as a function of the
adopted parameters.

The resulting maximum of the ${\rm FWHM}$ as a function of $\alpha$
or $\beta$  is shown in Fig.~\ref{fig:FWHM_limits}. All models
show clear dependence of the limits on the black hole mass (dashed and 
continuous lines for $10^6 M_{\odot}$ and $10^8 M_{\odot}$, correspondingly) 

Classical ADAF solution allows for very broad lines if the viscosity 
is small. Model B also displays large parameter range coming from 
adoption of weak or strong magnetic field. In the case of model C the
dependence of the FWHM is shallower and very broad lines are expected only
for very low viscosity {\it and} very weak magnetic field; otherwise,
evaporation is too efficient for a cold disk to survive close to a black
hole.

\section{Comparison with observations}

Theoretical constraints for the properties of the broad line objects can be
compared with the observed properties. We attempt to perform such a 
comparison using the sample of objects collected
from the literature. The measured quantities as well as estimated
black hole masses and Eddington rates for these objects are summarized in 
Table~\ref{tab1}. These objects, taken from various samples, do not
form an unbiased sample of AGN so the conclusions cannot be firm. Nevertheless,
such an analysis shows the effectiveness of the method and offers some
indicative results. 

\subsection{Data}

Optical/UV spectra of LBQS (large Bright Quasar Survey) objects included in 
the sample of Forster et al. (2001) were studied with MMT and du Point
telescopes. We
selected objects for which reliable MgII line measurements were performed, 
supplemented by the measurement of the continuum at 2800 \AA, rest frame.
The measured flux was corrected for the Galactic reddening.

The mass of the central black hole was calculated from the relation found by
Kaspi et al. (2000)
\begin{equation}
 M = 4.817  \left({\lambda L_\lambda (5100 \AA) \over 10^{44}} 
\right)^{0.7} FWHM^2~~~M_{\odot}.
\label{eq:masa}
\end{equation}

We determined the $\lambda L_\lambda (5100 \AA)$
from the measured flux at 2800 \AA~ by extrapolating the 
measured flux to 5100 \AA,
assuming a power law spectrum with a slope 0.5 
(i.e. $ F_{\nu} \propto \nu^{-0.5}$). The cosmological model 
used throughout the
paper is  $H_o = 75$ km s$^{-1}$ Mpc$^{-1}$ and $q_o = 0.5$. 
We used Eq.~\ref{eq:bolo} to determine the bolometric luminosity of an object,
and subsequently its Eddington ratio, $L_{bol}/L_{Edd}$, with $L_{Edd}$ 
defined by Eq.~\ref{eq:Edd}. 
The measured quantities (FWHM and $F_{\lambda}$), the derived 
black hole mass and the Eddington
ratio are given in Table~\ref{tab1}.

Similarly, for objects from the sample of Kuraszkiewicz et al. (2002) and 
Kuraszkiewicz et al. (2004) the FWHM of the MgII line was measured together
with the continuum at 2800 \AA, rest frame. The measurements were performed 
for HST data. The black hole mass of these sources and the Eddington ratio 
were derived as above.

For objects taken from Kaspi et al. (2000), as well as for those originally 
included in the study of Wandel et al. (1999), the measurements of FWHM of 
H$\beta$
lines are available, with the exception of PG~1351+640 for which an 
$H_{\alpha}$ 
measurement
was given, but corrected by 20 \% to account for a systematic difference 
between H$_{\alpha}$ and  H$_{\beta}$ line widths. 
The luminosity of the objects was 
provided in the form of $\lambda L_{\lambda}$ at 5100 \AA, assuming the Hubble
constant $H_o = 75$ km s$^{-1}$ Mpc$^{-1}$ and the deceleration parameter
$q_o = 0.5$. The values of the black hole masses for these objects were derived
from the reverberation measurements and were given by Kaspi et al. (2000),
and we determine the Eddington ratio using the bolometric 
luminosity estimate of
Eq.~\ref{eq:bolo}. 

For the nearby quasars taken from Shang et al. (2003) the 
FWHM of H$\beta$ was given 
together 
with $L_{\nu}(1549\AA)$. This last value was given for  $H_o = 50$ km s$^{-1}$ 
Mpc$^{-1}$ and $q_o = 0.5$ so it was necessary to recalculate it to  
$H_o = 75$ km s$^{-1}$ Mpc$^{-1}$ adopted in our paper. The black hole mass
and the Eddington ratio were obtained as for Forster at al. and 
Kuraszkiewicz et al. objects.

We also include several more distant quasars from Brotherton et al. (1994a).
We include in our sample only the radio-quiet sources.
The MgII widths were measured for those sources, and the continuum 
in magnitudes measured at 2200 \AA ~ was provided 
by Steidel \& Sargent (1991). 
The magnitude values were reversed into the flux measurement using
the definition of Steidel \& Sargent (1991), and subsequently the 
absolute flux at 2200 \AA~ was derived assuming the same
cosmological model as before. The black hole mass and the Eddington
ratio were determined as for Forster at al. and 
Kuraszkiewicz et al. objects.

\begin{deluxetable}{lccccrrc}
\tablewidth{0pt}
\tablenum{1}
\tablecaption{Broad line objects: measured line widths and luminosities, 
estimates of central black holes masses and Eddington ratios, as described
in the text, and the references}
\label{tab1}
\tablehead{
\colhead{Name} &
\colhead{Spectrum} &
\colhead{Redshift} &
\colhead{FWHM(MgII)} &
\colhead{$F_{\lambda}$(2800 \AA)} &
\colhead{log M} &
\colhead{log $L/L_{Edd}$} &
\colhead{Ref}
}


\startdata
LBQS 0010+0146 & 0010+0146la  &   0.587 &    13000. &     $-$15.3856 &     9.72 &    $-$1.58 &   1\\ 
LBQS 0018$-$0252 & 0018$-$0252la  &   0.618 &     5600. &     $-$15.4883 &     8.96 &    $-$0.86 &   1\\ 
LBQS 0022$-$0140 & 0022$-$0140la  &   0.776 &     8500. &     $-$15.4825 &     9.51 &    $-$1.14 &   1\\ 
LBQS 0028$-$0101 & 0028$-$0101la  &   0.543 &    12000. &     $-$15.8062 &     9.29 &    $-$1.67 &   1\\ 
LBQS 0042$-$2750 & 0042$-$2750la  &   0.741 &     6400. &     $-$15.7581 &     9.04 &    $-$0.99 &   1\\ 
LBQS 0047$-$3059 & 0047$-$3059la  &   0.559 &     8000. &     $-$14.5679 &     9.83 &    $-$0.93 &   1\\ 
LBQS 0048$-$0133 & 0048$-$0133la  &   0.763 &     8500. &     $-$14.5792 &    10.13 &    $-$0.88 &   1\\ 
LBQS 0055$-$2948 & 0055$-$2948la  &   0.663 &     3940. &     $-$15.7641 &     8.52 &    $-$0.61 &   1\\ 
LBQS 0056$-$0009 & 0056$-$0009la  &   0.717 &     7000. &     $-$14.7753 &     9.77 &    $-$0.79 &   1\\ 
LBQS 0057+0000 & 0057+0000la  &   0.776 &     8000. &     $-$14.7906 &     9.95 &    $-$0.88 &   1\\ 
LBQS 0102$-$2713 & 0102$-$2713la  &   0.780 &     5400. &     $-$15.5035 &     9.11 &    $-$0.75 &   1\\ 
LBQS 0102$-$0147 & 0102$-$0147la  &   0.571 &    10000. &     $-$14.9600 &     9.76 &    $-$1.24 &   1\\ 
LBQS 0103$-$2622 & 0103$-$2622la  &   0.776 &     7000. &     $-$15.4250 &     9.39 &    $-$0.95 &   1\\ 
LBQS 0107$-$0235 & 0107$-$0235la  &   0.958 &    10500. &     $-$15.3280 &     9.99 &    $-$1.20 &   1\\ 
LBQS 0252+0141 & 0252+0141la  &   0.621 &    13000. &     $-$15.1197 &     9.95 &    $-$1.48 &   1\\ 
LBQS 0253$-$0138 & 0253$-$0138la  &   0.879 &    16500. &     $-$14.9133 &    10.60 &    $-$1.50 &   1\\ 
LBQS 0257+0025 & 0257+0025la  &   0.532 &     6400. &     $-$14.7526 &     9.46 &    $-$0.81 &   1\\ 
LBQS 0301+0010 & 0301+0010la  &   0.638 &     6600. &     $-$14.7610 &     9.63 &    $-$0.78 &   1\\ 
LBQS 0303$-$0132 & 0303$-$0132la  &   0.606 &     6500. &     $-$14.7173 &     9.61 &    $-$0.77 &   1\\ 
LBQS 0303$-$0241 & 0303$-$0241la  &   0.686 &     5000. &     $-$15.0694 &     9.24 &    $-$0.60 &   1\\ 
LBQS 0305+0222 & 0305+0222la  &   0.590 &    10400. &     $-$14.8150 &     9.93 &    $-$1.21 &   1\\ 
LBQS 0307$-$0015 & 0307$-$0015la  &   0.770 &     6400. &     $-$14.8683 &     9.69 &    $-$0.71 &   1\\ 
LBQS 1027$-$0149 & 1027$-$0149la  &   0.754 &    11500. &     $-$15.5531 &     9.70 &    $-$1.43 &   1\\ 
LBQS 1137+0051 & 1137+0051la  &   0.874 &     9500. &     $-$15.8626 &     9.45 &    $-$1.31 &   1\\ 
LBQS 1138+0003 & 1138+0003la  &   0.500 &    14000. &     $-$15.4346 &     9.61 &    $-$1.72 &   1\\ 
LBQS 1205+1729 & 1205+1729la  &   0.548 &    16000. &     $-$15.0968 &    10.04 &    $-$1.70 &   1\\ 
LBQS 1211+0841 & 1211+0841la  &   0.585 &     8800. &     $-$15.3781 &     9.38 &    $-$1.24 &   1\\ 
LBQS 1222+1640 & 1222+1640la  &   0.549 &     9600. &     $-$15.0785 &     9.61 &    $-$1.25 &   1\\ 
LBQS 1222+0901 & 1222+0901la  &   0.535 &    13000. &     $-$13.4831 &    10.97 &    $-$1.04 &   1\\ 
LBQS 1224+1604 & 1224+1604la  &   0.533 &     7500. &     $-$15.6004 &     9.01 &    $-$1.20 &   1\\ 
LBQS 1236+1802 & 1236+1802la  &   0.517 &    14000. &     $-$15.4997 &     9.60 &    $-$1.72 &   1\\ 
LBQS 1237+0950 & 1237+0950la  &   0.736 &    11000. &     $-$15.1948 &     9.89 &    $-$1.30 &   1\\ 
LBQS 1240+0224 & 1240+0224la  &   0.790 &    13000. &     $-$15.3358 &    10.00 &    $-$1.46 &   1\\ 
LBQS 1240+1745 & 1240+1745la  &   0.549 &    11500. &     $-$14.9485 &     9.86 &    $-$1.37 &   1\\ 
LBQS 1243+1456 & 1243+1456la  &   0.582 &    10500. &     $-$15.3523 &     9.55 &    $-$1.39 &   1\\ 
LBQS 1244+1329 & 1244+1329la  &   0.512 &     2600. &     $-$14.7649 &     8.64 &    $-$0.05 &   1\\ 
LBQS 1245+1719 & 1245+1719la  &   0.752 &    10000. &     $-$15.2126 &     9.82 &    $-$1.21 &   1\\ 
LBQS 1250+0109 & 1250+0109la  &   0.792 &    13200. &     $-$15.6707 &     9.78 &    $-$1.57 &   1\\ 
LBQS 1313$-$0228 & 1313$-$0228la  &   0.704 &     5200. &     $-$15.6822 &     8.86 &    $-$0.81 &   1\\ 
LBQS 1315+0150 & 1315+0150la  &   0.539 &    12500. &     $-$15.0981 &     9.81 &    $-$1.49 &   1\\ 
LBQS 1319+0033 & 1319+0033la  &   0.530 &    11400. &     $-$15.2236 &     9.63 &    $-$1.45 &   1\\ 
LBQS 1322$-$0204 & 1322$-$0204la  &   0.573 &     7800. &     $-$14.9842 &     9.53 &    $-$1.03 &   1\\ 
LBQS 1323+0205 & 1323+0205la  &   0.641 &     4400. &     $-$15.4965 &     8.77 &    $-$0.64 &   1\\ 
LBQS 1326+0206 & 1326+0206la  &   1.442 &    13000. &     $-$15.1133 &    10.70 &    $-$1.16 &   1\\ 
LBQS 1326$-$0257 & 1326$-$0257la  &   0.743 &     5000. &     $-$15.3303 &     9.12 &    $-$0.65 &   1\\ 
LBQS 1328+0205 & 1328+0205la  &   0.692 &     3500. &     $-$14.3684 &     9.43 &    $-$0.08 &   1\\ 
LBQS 1340$-$0020 & 1340$-$0020la  &   0.786 &    11000. &     $-$15.6413 &     9.64 &    $-$1.41 &   1\\ 
LBQS 1433+0011 & 1433+0011la  &   0.583 &     9000. &     $-$15.4353 &     9.36 &    $-$1.28 &   1\\ 
LBQS 1440$-$0234 & 1440$-$0234la  &   0.678 &     4400. &     $-$14.2734 &     9.67 &    $-$0.26 &   1\\ 
LBQS 1445+0222 & 1445+0222la  &   0.775 &     6800. &     $-$15.3117 &     9.44 &    $-$0.90 &   1\\ 
LBQS 2132$-$4228 & 2132$-$4228la  &   0.569 &     6500. &     $-$15.9061 &     8.72 &    $-$1.15 &   1\\ 
LBQS 2142$-$4318 & 2142$-$4318la  &   1.118 &     5500. &     $-$16.0346 &     9.07 &    $-$0.79 &   1\\ 
LBQS 2154$-$2105 & 2154$-$2105la  &   0.575 &     6500. &     $-$14.7796 &     9.52 &    $-$0.81 &   1\\ 
LBQS 2203$-$2134 & 2203$-$2134la  &   0.576 &    13500. &     $-$15.7686 &     9.46 &    $-$1.74 &   1\\ 
LBQS 2244+0020 & 2244+0020la  &   0.973 &    12500. &     $-$14.7397 &    10.57 &    $-$1.17 &   1\\ 
LBQS 2350$-$0012 & 2350$-$0012la  &   0.561 &     5600. &     $-$14.6781 &     9.44 &    $-$0.65 &   1\\ 
LBQS 2353+0032 & 2353+0032la  &   0.558 &    12500. &     $-$15.4269 &     9.61 &    $-$1.58 &   1\\ 
  \hline                                                               
&&&&&&&\\
HST/FOS PRE-COSTAR:\\                                                   
  \hline                                                               
 III ZW 2   &    0010+1058ra  &   0.089 &     8700. &     $-$14.1559 &     8.86 &    $-$1.45 &   2\\ 
 I ZW 1     &    0053+1241ra  &   0.061 &     5550. &     $-$13.7892 &     8.48 &    $-$1.06 &   2\\ 
 Fairall 9  &    0123$-$5848ra  &   0.047 &     5750. &     $-$13.7382 &     8.38 &    $-$1.14 &   2\\ 
 PKS 0403$-$13&    0405$-$1308ra  &   0.571 &     8600. &     $-$14.9150 &     9.66 &    $-$1.09 &   2\\ 
 3C 207     &    0840+1312ra  &   0.681 &     9500. &     $-$15.1675 &     9.72 &    $-$1.19 &   2\\ 
 3C 215     &    0906+1646ra  &   0.412 &    14400. &     $-$15.2551 &     9.61 &    $-$1.75 &   2\\ 
 B2 0923+392&    0927+3902ra  &   0.695 &    12000. &     $-$14.6144 &    10.33 &    $-$1.22 &   2\\ 
 B2 0923+392&    0927+3902rb  &   0.695 &    13900. &     $-$14.8952 &    10.26 &    $-$1.43 &   2\\ 
 3C 263     &    1139+6547ra  &   0.646 &    17000. &     $-$14.5002 &    10.65 &    $-$1.51 &   2\\ 
 PG 1211+143&    1214+1403ra  &   0.081 &     5400. &     $-$13.8469 &     8.60 &    $-$0.97 &   2\\ 
 MRK 205    &    1221+7518rb  &   0.071 &     6850. &     $-$13.9038 &     8.68 &    $-$1.23 &   2\\ 
 PG 1226+023&    1229+0203ra  &   0.158 &     7400. &     $-$13.0137 &     9.90 &    $-$0.80 &   2\\ 
 3C 277.1   &    1252+5634ra  &   0.321 &     7100. &     $-$15.0441 &     8.95 &    $-$1.16 &   2\\ 
 3C 279     &    1256$-$0547ra  &   0.536 &    17000. &     $-$14.7935 &    10.29 &    $-$1.67 &   2\\ 
 NGC 5548   &    1417+2508ra  &   0.017 &     8450. &     $-$14.2071 &     7.75 &    $-$1.89 &   2\\ 
 PG 1415+451&    1417+4456ra  &   0.114 &     4500. &     $-$14.4230 &     8.26 &    $-$0.89 &   2\\ 
 3C 334     &    1620+1736rb  &   0.555 &     9600. &     $-$14.5985 &     9.96 &    $-$1.10 &   2\\ 
 3C 345     &    1642+3948ra  &   0.593 &    10000. &     $-$14.6169 &    10.03 &    $-$1.12 &   2\\ 
B2 2201+315A&    2203+3145ra  &   0.295 &     8200. &     $-$13.9539 &     9.78 &    $-$0.98 &   2\\ 
 4C 11.72   &    2254+1136ra  &   0.326 &     9000. &     $-$14.2875 &     9.70 &    $-$1.13 &   2\\ 
  \hline                                                               
&&&&&&&\\
HST/FOS POST-COSTAR: \\                                                 
\hline
 MRK 335       & 0006+2012oa  &   0.026 &     5300. &     $-$13.4173 &     8.16 &    $-$1.14 &   3\\ 
 PG 0026+129   & 0029+1316ob  &   0.142 &     6700. &     $-$13.9626 &     9.08 &    $-$1.03 &   3\\ 
 MRK 348       & 0048+3157oa  &   0.015 &     2400. &     $-$15.8095 &     5.45 &    $-$1.31 &   3\\ 
 I ZW 1        & 0053+1241oa  &   0.061 &     5850. &     $-$13.7184 &     8.57 &    $-$1.08 &   3\\ 
 0318$-$196      & 0320$-$1926oa  &   0.104 &    10000. &     $-$15.7505 &     7.96 &    $-$2.01 &   3\\ 
 AKN 120       & 0516$-$0008oa  &   0.033 &     9100. &     $-$13.2105 &     8.92 &    $-$1.48 &   3\\ 
 PG 0947+396   & 0950+3926oa  &   0.206 &     5300. &     $-$14.3277 &     8.88 &    $-$0.83 &   3\\ 
 PG 1001+054   & 1004+0513oa  &   0.161 &     4800. &     $-$14.6082 &     8.42 &    $-$0.90 &   3\\ 
 B2 1028+313   & 1030+3102oa  &   0.178 &    10000. &     $-$14.5558 &     9.17 &    $-$1.49 &   3\\ 
 ZW 212.025    & 1034+3938oa  &   0.042 &     3100. &     $-$14.9705 &     6.91 &    $-$1.01 &   3\\ 
 NGC 3516      & 1106+7234oa  &   0.009 &     6450. &     $-$13.4862 &     7.63 &    $-$1.61 &   3\\ 
 NGC 3516      & 1106+7234ob  &   0.009 &     6000. &     $-$13.5925 &     7.49 &    $-$1.58 &   3\\ 
 NGC 3516      & 1106+7234oc  &   0.009 &     6650. &     $-$13.6224 &     7.56 &    $-$1.68 &   3\\ 
 NGC 3516      & 1106+7234od  &   0.009 &     6000. &     $-$13.4524 &     7.59 &    $-$1.53 &   3\\ 
 NGC 3516      & 1106+7234oe  &   0.009 &     7550. &     $-$13.9228 &     7.46 &    $-$1.88 &   3\\ 
 PG 1114+445   & 1117+4413oa  &   0.144 &     7000. &     $-$14.3122 &     8.88 &    $-$1.17 &   3\\ 
 PG 1114+445   & 1117+4413ob  &   0.144 &     6600. &     $-$14.4533 &     8.73 &    $-$1.16 &   3\\ 
 MC 1118+12    & 1121+1236oa  &   0.685 &     9500. &     $-$15.4223 &     9.55 &    $-$1.26 &   3\\ 
 GQ COMAE      & 1204+2754oa  &   0.165 &    12000. &     $-$14.8342 &     9.08 &    $-$1.76 &   3\\ 
 1219+047      & 1221+0430ob  &   0.094 &    10000. &     $-$15.1540 &     8.31 &    $-$1.86 &   3\\ 
 QSO1219+057   & 1219+0545oa  &   0.114 &     3300. &     $-$15.5099 &     7.23 &    $-$0.95 &   3\\ 
 1220+1601     & 1223+1545ob  &   0.081 &     6500. &     $-$16.1004 &     7.18 &    $-$1.81 &   3\\ 
 NGC 4579      & 1237+1149oa  &   0.005 &     6800. &     $-$15.3764 &     5.99 &    $-$2.38 &   3\\ 
 PG 1322+659   & 1323+6541oa  &   0.168 &    10000. &     $-$14.3634 &     9.26 &    $-$1.45 &   3\\ 
 MRK 270       & 1341+6740oa  &   0.009 &     2000. &     $-$15.9921 &     4.85 &    $-$1.34 &   3\\ 
 PG 1352+183   & 1354+1805oa  &   0.152 &     5200. &     $-$14.3633 &     8.62 &    $-$0.91 &   3\\ 
 PG 1415+451   & 1417+4456oa  &   0.114 &     5900. &     $-$14.3667 &     8.53 &    $-$1.11 &   3\\ 
 MRK 478       & 1442+3526oa  &   0.079 &     6400. &     $-$13.8295 &     8.74 &    $-$1.12 &   3\\ 
 MR 2251$-$178   & 2254$-$1734oa  &   0.068 &    10950. &     $-$13.9389 &     9.03 &    $-$1.66 &   3\\ 
 NGC 7469      & 2303+0852oa  &   0.016 &     4200. &     $-$13.5032 &     7.59 &    $-$1.09 &   3\\ 
 IRAS13349+2438& 1337+2423ob  &   0.108 &     9300. &     $-$14.7569 &     8.62 &    $-$1.64 &   3\\ 
 ARP 102B      & 1719+4858oa  &   0.024 &    15500. &     $-$15.2975 &     7.72 &    $-$2.65 &   3\\ 
 CYGNUS A      & 1959+4044oa  &   0.056 &    11500. &     $-$15.1480 &     8.10 &    $-$2.12 &   3\\ 
  \hline 
&&&&&&&\\                                                              
 Name & &&FWHM($H_{\beta}$) & $\lambda F_{\lambda}$(5100 \AA)&log M & log $L/L_{Edd}$ & Ref.\\                        
 \hline  
 3C 120    &\ldots &\ldots  &     1910. &         0.73000 &     7.36 &    $-$0.64 &   4\\ 
 3C 390.3  &\ldots&\ldots  &    10000. &         0.64000 &     8.53 &    $-$1.87 &   4\\ 
 Akn 120   &\ldots&\ldots  &     5800. &         1.39000 &     8.26 &    $-$1.27 &   4\\ 
 F9        &\ldots&\ldots  &     5780. &         1.37000 &     7.90 &    $-$0.91 &   4\\ 
 IC 4329A  &\ldots&\ldots  &     5050. &         0.16400 &     6.70 &    $-$0.63 &   4\\ 
 Mrk 79    &\ldots&\ldots  &     4470. &         0.42300 &     7.72 &    $-$1.24 &   4\\ 
 Mrk 110   &\ldots&\ldots  &     1430. &         0.38000 &     6.75 &    $-$0.31 &   4\\ 
 Mrk 335   &\ldots&\ldots  &     1620. &         0.62200 &     6.80 &    $-$0.15 &   4\\ 
 Mrk 509   &\ldots&\ldots  &     2270. &         1.47000 &     7.76 &    $-$0.74 &   4\\ 
 Mrk 590   &\ldots&\ldots  &     2470. &         0.51000 &     7.25 &    $-$0.69 &   4\\ 
 Mrk 817   &\ldots&\ldots  &     4490. &         0.52600 &     7.64 &    $-$1.07 &   4\\ 
 NGC 3227  &\ldots&\ldots  &     4920. &         0.02020 &     7.59 &    $-$2.43 &   4\\ 
 NGC 3783  &\ldots&\ldots  &     3790. &         0.17700 &     6.97 &    $-$0.87 &   4\\ 
 NGC 4051  &\ldots&\ldots  &     1170. &         0.00525 &     6.11 &    $-$1.54 &   4\\ 
 NGC 4151  &\ldots&\ldots  &     5910. &         0.07200 &     7.18 &    $-$1.47 &   4\\ 
 NGC 5548  &\ldots&\ldots  &     6300. &         0.27000 &     8.09 &    $-$1.80 &   4\\ 
 NGC 7469  &\ldots&\ldots  &     3000. &         0.55300 &     6.81 &    $-$0.22 &   4\\ 
  \hline                                                               
&&&&&&&\\
 Name & &&FWHM($H_{\beta}$) & $\lambda F_{\lambda}$(5100 \AA)&log M & log $L/L_{Edd}$ & Ref.\\                        
  \hline                                                               
 PG 0026+129   &\ldots&\ldots  &     2100. &         7.00000 &     7.73 &    $-$0.03 &   5\\ 
 PG 0052+251   &\ldots&\ldots  &     3990. &         6.50000 &     8.34 &    $-$0.68 &   5\\ 
 PG 0804+761   &\ldots&\ldots  &     2984. &         6.60000 &     8.28 &    $-$0.60 &   5\\ 
 PG 0844+349   &\ldots&\ldots  &     2730. &         1.72000 &     7.33 &    $-$0.25 &   5\\ 
 PG 0953+414   &\ldots&\ldots  &     2885. &        11.90000 &     8.26 &    $-$0.34 &   5\\ 
 PG 1211+143   &\ldots&\ldots  &     1832. &         4.93000 &     7.61 &    $-$0.06 &   5\\ 
 PG 1226+023   &\ldots&\ldots  &     3416. &        64.40000 &     8.74 &    $-$0.08 &   5\\ 
 PG 1229+204   &\ldots&\ldots  &     3440. &         0.94000 &     7.88 &    $-$1.05 &   5\\ 
 PG 1307+085   &\ldots&\ldots  &     4190. &         5.27000 &     8.45 &    $-$0.87 &   5\\ 
 PG 1351+640   &\ldots&\ldots  &     1404. &         4.38000 &     7.66 &    $-$0.17 &   5\\ 
 PG 1411+442   &\ldots&\ldots  &     2456. &         3.25000 &     7.90 &    $-$0.54 &   5\\ 
 PG 1426+015   &\ldots&\ldots  &     6250. &         4.09000 &     8.67 &    $-$1.21 &   5\\ 
 PG 1613+658   &\ldots&\ldots  &     7000. &         6.69000 &     8.38 &    $-$0.70 &   5\\ 
 PG 1617+175   &\ldots&\ldots  &     5120. &         2.37000 &     8.44 &    $-$1.21 &   5\\ 
 PG 1700+518   &\ldots&\ldots  &     2180. &        27.10000 &     7.78 &     0.51 &   5\\ 
 PG 1704+608   &\ldots&\ldots  &      890. &        35.60000 &     7.57 &     0.84 &   5\\ 
 PG 2130+099   &\ldots&\ldots  &     2410. &         2.16000 &     8.16 &    $-$0.97 &   5\\ 
  \hline                                                               
&&&&&&&\\
 Name  & &Redshift & FWHM($H_{\beta}$) & log $L_{\nu}$(1549 \AA) & log M &  log $L/L_{Edd}$ & Ref. \\                           
  \hline                                                               
 PG 0947+396 &\ldots&   0.2056 &     4830. &    30.33 &     8.76 &    $-$0.89 &   6\\ 
 PG 0953+414 &\ldots&   0.2341 &     3130. &    30.59 &     8.56 &    $-$0.44 &   6\\ 
 PG 1001+054 &\ldots&   0.1603 &     1740. &    29.73 &     7.45 &    $-$0.19 &   6\\ 
 PG 1114+445 &\ldots&   0.1440 &     4570. &    29.80 &     8.34 &    $-$1.01 &   6\\ 
 PG 1115+407 &\ldots&   0.1541 &     1720. &    30.08 &     7.68 &    $-$0.07 &   6\\ 
 PG 1116+215 &\ldots&   0.1759 &     2920. &    30.67 &     8.56 &    $-$0.36 &   6\\ 
 PG 1202+281 &\ldots&   0.1651 &     5050. &    29.51 &     8.22 &    $-$1.18 &   6\\ 
 PG 1216+069 &\ldots&   0.3319 &     5190. &    30.69 &     9.07 &    $-$0.85 &   6\\ 
 PG 1226+023 &\ldots&   0.1575 &     3520. &    31.37 &     9.21 &    $-$0.31 &   6\\ 
 PG 1309+355 &\ldots&   0.1823 &     2940. &    30.21 &     8.24 &    $-$0.50 &   6\\ 
 PG 1322+659 &\ldots&   0.1675 &     2790. &    30.02 &     8.06 &    $-$0.51 &   6\\ 
 PG 1352+183 &\ldots&   0.1510 &     3600. &    30.04 &     8.30 &    $-$0.73 &   6\\ 
 PG 1402+261 &\ldots&   0.1650 &     1910. &    30.40 &     8.00 &    $-$0.07 &   6\\ 
 PG 1411+442 &\ldots&   0.0895 &     2670. &    29.56 &     7.70 &    $-$0.61 &   6\\ 
 PG 1415+451 &\ldots&   0.1143 &     2620. &    29.72 &     7.80 &    $-$0.55 &   6\\ 
 PG 1425+267 &\ldots&   0.3637 &     9410. &    30.42 &     9.40 &    $-$1.45 &   6\\ 
 PG 1427+480 &\ldots&   0.2203 &     2540. &    30.20 &     8.11 &    $-$0.38 &   6\\ 
 PG 1440+356 &\ldots&   0.0773 &     1450. &    29.90 &     7.41 &     0.02 &   6\\ 
 PG 1444+407 &\ldots&   0.2670 &     2480. &    30.63 &     8.39 &    $-$0.23 &   6\\ 
 PG 1512+370 &\ldots&   0.3700 &     6810. &    30.66 &     9.28 &    $-$1.09 &   6\\ 
 PG 1543+489 &\ldots&   0.4000 &     1560. &    30.75 &     8.07 &     0.21 &   6\\ 
 PG 1626+554 &\ldots&   0.1317 &     4490. &    30.11 &     8.54 &    $-$0.90 &   6\\ 
   \hline                                                              
&&&&&&&\\
 Name & &Redshift & FWHM(MgII) &   M (2200 \AA) & log M &  log $L/L_{Edd}$ & Ref. \\                       
   \hline                                                              
 0009-0138  &\ldots&  2.0000 &     3500. &     -28.0 &    10.04 &     0.24 &   7\\ 
 0013-0029  &\ldots&  2.0860 &     2200. &     -27.0 &     9.37 &     0.53 &   7\\ 
 0019+0107  &\ldots&  2.1340 &     5700. &     -28.0 &    10.48 &    -0.18 &   7\\ 
 0058+0155  &\ldots&  1.9590 &     4200. &     -28.0 &    10.19 &     0.08 &   7\\ 
 0150-2015  &\ldots&  2.1470 &     2800. &     -28.0 &     9.86 &     0.44 &   7\\ 
 0206+0008  &\ldots&  1.8900 &     4100. &     -28.0 &    10.16 &     0.09 &   7\\ 
 0348+0610  &\ldots&  2.0600 &     3400. &     -29.0 &    10.30 &     0.39 &   7\\ 
 0848+1623  &\ldots&  2.0090 &     6100. &     -28.0 &    10.52 &    -0.24 &   7\\ 
 0856+4649  &\ldots&  0.9240 &     6400. &     -27.0 &    10.08 &    -0.49 &   7\\ 
 0920+5800  &\ldots&  1.3760 &     3600. &     -27.0 &     9.68 &     0.05 &   7\\ 
 0932+5006  &\ldots&  1.8800 &     6500. &     -28.0 &    10.56 &    -0.31 &   7\\ 
 0946+3009  &\ldots&  1.2160 &     3700. &     -28.0 &     9.95 &     0.13 &   7\\ 
 0957+5542  &\ldots&  2.1000 &     2500. &     -28.0 &     9.76 &     0.54 &   7\\ 
 1206+4557  &\ldots&  1.1580 &     4300. &     -28.0 &    10.07 &     0.00 &   7\\ 
 1222+0223  &\ldots&  2.0500 &     3100. &     -28.0 &     9.94 &     0.35 &   7\\ 
 1228+0750  &\ldots&  1.8130 &     2400. &     -28.0 &     9.68 &     0.55 &   7\\ 
 1245+3431  &\ldots&  2.0680 &     2800. &     -27.0 &     9.57 &     0.31 &   7\\ 
 1246+3746  &\ldots&  2.2120 &     5500. &     -27.0 &    10.18 &    -0.26 &   7\\ 
 1248+4007  &\ldots&  1.3030 &     3400. &     -27.0 &     9.62 &     0.09 &   7\\ 
 1254+0443  &\ldots&  1.0240 &     8300. &     -27.0 &    10.33 &    -0.71 &   7\\ 
 1317+5203  &\ldots&  1.0220 &     4700. &     -27.0 &     9.84 &    -0.21 &   7\\ 
 1321+2925  &\ldots&  1.3300 &     5000. &     -26.0 &     9.68 &    -0.36 &   7\\ 
 1338+4138  &\ldots&  1.2190 &     3900. &     -28.0 &    10.00 &     0.09 &   7\\ 
 1407+2632  &\ldots&  0.9440 &     8200. &     -27.0 &    10.30 &    -0.71 &   7\\ 
 1416+0906  &\ldots&  2.0000 &     2800. &     -28.0 &     9.84 &     0.43 &   7\\ 
 1421+3305  &\ldots&  1.9040 &     2700. &     -28.0 &     9.80 &     0.46 &   7\\ 
 1522+1009  &\ldots&  1.3210 &     6000. &     -28.0 &    10.39 &    -0.28 &   7\\ 
 1634+7037  &\ldots&  1.3340 &     3400. &     -29.0 &    10.18 &     0.34 &   7\\ 
 1704+7101  &\ldots&  2.0000 &     4500. &     -28.0 &    10.26 &     0.02 &   7\\ 
 1715+5331  &\ldots&  1.9290 &     5000. &     -29.0 &    10.62 &     0.04 &   7\\ 
 1718+4807  &\ldots&  1.0840 &     2600. &     -28.0 &     9.62 &     0.43 &   7\\ 
 2302+0255  &\ldots&  1.0440 &     7100. &     -27.0 &    10.20 &    -0.57 &   7\\ 
 2340-0019  &\ldots&  0.8960 &     4000. &     -27.0 &     9.67 &    -0.09 &   7\\  
\enddata
\tablecomments{
{\bf References: 1.} Forster et al. (2002)
            {\bf 2.} Kuraszkiewicz et al. 2002
	    {\bf 3.} Kuraszkiewicz et al. 2004
	    {\bf 4.} Kaspi et al. (2000), after Wandel et al. (1999)
	    {\bf 5.} Kaspi et al. (2000)
	    {\bf 6.} Shang et al. (2003)
	    {\bf 7.} Brotherton et al. (1994a).
All luminosities are for $H_o = 75$ km s$^{-1}$ Mpc$^{-1}$ and $q_o = 0.5$. 
}

\end{deluxetable}

\begin{figure}
\epsfxsize = 80 mm \epsfbox[50 180 560 1530]{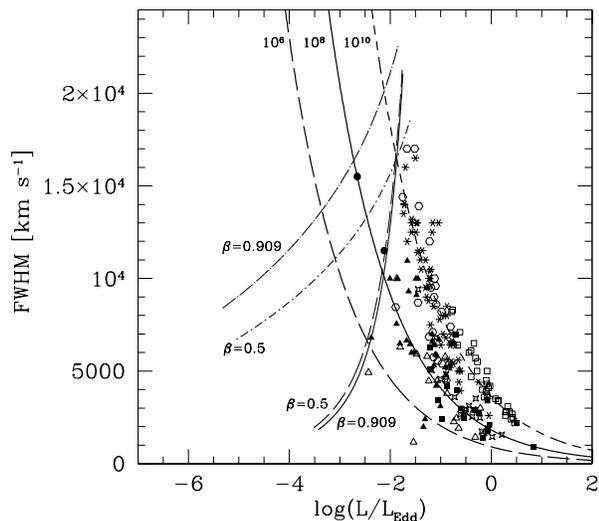}
\caption{The dependence of full width of half maximum of the broad line 
on bolometric luminosity. 
Observational points come from Table~\ref{tab1}.
Lines marked with the corresponding values of the black hole mass 
show the trends given by Eq.~\ref{eq:FWHM}. Other lines constrain
the parameter space of the Broad Line objects:
continuous line and long-dash line represent the ADAF (model A) with
the viscosity $\alpha = 0.04$ in the case
of negligible magnetic field and for $\beta = 0.5$, correspondingly;
dashed lines show the limits based on model C assuming 
viscosity parameter $\alpha=0.02$ and  
again two values of $\beta$ (0.909 and 0.5, correspondingly).
}
\label{fig:data1}
\end{figure}

\begin{figure}
\epsfxsize = 80 mm \epsfbox[50 180 560 660]{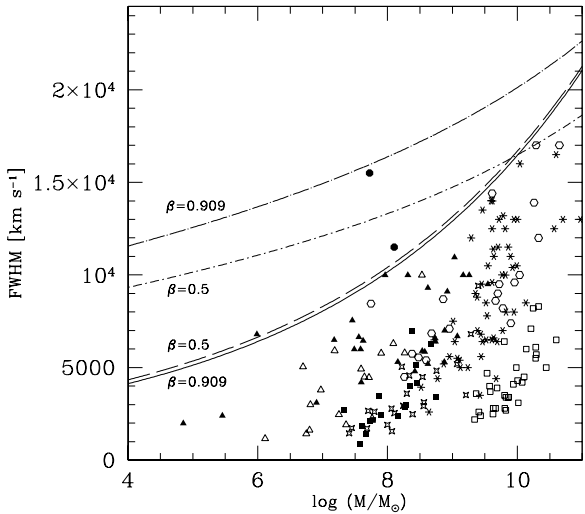}
\caption{The dependence of full width of half maximum on the black hole mass. 
Observational points come from Table~\ref{tab1}.
Continuous and long-dashed lines represent the limits 
based on model A, viscosity parameter 
$\alpha=0.04$, $\beta = 0.909$ and $\beta = 0.5$, correspondingly.  
Short-dash-dot line and long-dash-dot lines are 
for $\beta= 0.5$ and $\beta=0.909$.
}
\label{fig:data2}
\end{figure}

\subsection{Results} 

The Eddington ratios of practically all objects from Table~\ref{tab1} are
larger than 0.01. All these objects are classified as typical Broad Line 
objects: QSO, Seyfert 1 galaxies or Narrow Line Seyfert 1 galaxies. 
The maximum values of the FWHM among the objects in our sample do not 
exceed 17 000 km/s, well within the conservative upper limit for a Broad Line
object adopted by Laor (2003b) in his discussion.
Therefore, our sample supports the conclusion 
that Broad Line objects exist above a certain threshold in the 
Eddington ratio, and the widths of their broad lines are limited.
Such an interpretation is consistent with an accretion scenario based on the 
evaporation of the inner cold disk in objects with a low Eddington ratio.

In order to determine which of the three specific models 
(see Section~\ref{sect:evap}) 
better represents the allowed parameter space for the Broad Line objects
we use two diagrams.  
In Fig.~\ref{fig:data1} we show the dependence between the line width and
the Eddington ratio for objects from our sample. Objects coming from various
subsamples are marked with different symbols but they all form a mostly uniform
distribution. 
The black dot marking an object with a rather large FWHM (15 000 km/s) and 
low Eddington ratio (0.0022) represents the source Arp 102B (the second to 
last source from Reference 3 in Table~\ref{tab1}), which belongs to the
class of radio loud objects with double peaked broad emission lines (Chen
\& Halpern 1989).

We did not plot error bars in Fig.~\ref{fig:data1} 
since they are rather difficult to
asses. The measurements of the FWHM are performed with 20-30 percent 
accuracy; the Eddington ratio is in most cases derived on the basis of
statistical correlations like those given by Eqs.~\ref{eq:bolo} 
and \ref{eq:masa} which show a certain intrinsic scatter.

The objects basically follow the relation between the line width and the 
Eddington ratio given by Eq.~\ref{eq:FWHM} marked in Fig.~\ref{fig:data1} for
three specific values of the black hole mass. Therefore, the relation found 
by Kaspi et al. (2000) seems to describe the whole sample of objects quite 
well.  The upper limits for the line width at a given Eddington ratio, 
estimated from two models, are marked as lines roughly perpendicular to 
the Kaspi et al. relations. A successful model should clearly separate 
the region occupied by objects from the avoidance region.

We plot four limiting lines coming from two models. The ADAF solution
(model A) is represented by two lines:
a continuous line for a negligible magnetic field and long dashed line for 
$\beta = 0.5$; 
$\alpha = 0.04$ in both cases. Evaporation model C is represented by
two other lines: a long-dash-dot line (negligible magnetic field) and a
short-dash-dot line  ($\beta = 0.5)$; $\alpha = 0.02$ in both cases.

Here it is apparent 
that all theoretical constraints for the line width are not 
universal, but show some dependence on the Eddington ratio: the line is
allowed to be broader for an object with a lower mass/lower Eddington ratio.
However, the strength of this coupling depends on the model. In the ADAF model
the limiting line is quite steep and the upper limit from the line width
is almost universal while in the evaporation model C the line is flatter
and the predicted limiting value for the line width depends considerably on 
the object mass and Eddington ratio. Meyer-Hofmeister \& Meyer (2001) have
shown some calculations of evaporation rate for different values of the 
viscosity parameter. They took into account a wind from the corona, which
can affect the general solution. The effect of incorporation of the wind
losses also depends on viscosity. Taking into account these results, 
it is apparent 
that the dependence of their model on viscosity is intermediate  between the 
ADAF and our evaporation model C. 

A change in the viscosity parameter results in essentially a 
parallel shift of a line,
without significant change of the line slope. For example, increasing
the viscosity parameter $\alpha$ results in moving 
the limiting line given by the 
model C closer to the data points but generally either some of the objects
with the broadest lines are already above the threshold or there is a large 
region to the left of the line which should be populated by broad line
objects but no data points are actually there. The steeper relation coming
from ADAF seems to better represent the populated region. The best value
of the viscosity parameter is around 0.04, as adopted in 
Fig.~\ref{fig:data1}.   

Similarly, we can assess the dependence of the line width on the black hole
mass. In Fig.~\ref{fig:data2} we plot the objects from our sample. Objects
occupy predominantly the lower right corner of the diagram. The allowed region,
according to a specific model, is shown with lines, as in Fig~\ref{fig:data1}.
Limits for the line width in both models depend on the mass 
(objects with larger masses are allowed to have broader lines) but the exact
shape of this dependence is different in the ADAF model and in the evaporation
model. Changing the viscosity parameter $\alpha$ it is possible to 
move the lines
up or down but without a strong change of the shape. Again it is apparent 
that the region populated by the objects from Table~\ref{tab1} is better
described by the classical ADAF than a more complex evaporation model, and the
favored viscosity parameter $\alpha = 0.04$.

\section{Discussion}

Results presented in this paper support the existence of a
connection between the BLR and the survival of the cold disk sufficiently
close to the central black hole. This agrees with
general conclusions of Nicastro et al. (2003) and Laor (2003b), but it takes  
this idea one step further.

Each specific mechanism of cold disk disruption/evaporation gives a specific
shape of the allowed region for broad line objects in the line width-Eddington 
ratio and line width-black hole mass diagram. We considered three models
of the inner cold disk disappearance and for two of them (classical ADAF
and generalized disk evaporation model of \Agata \& Czerny 2000b) we 
constructed the 
detailed theoretical region boundaries for the broad line region objects
in these two diagrams. We placed on this diagram a large sample of 
broad line AGN taken from the literature.

Perhaps surprisingly, the classical ADAF model represented the
disk disappearance conditions better than the disk evaporation model. 
The value of the viscosity 
best representing the allowed parameter space ($\alpha = 0.04$) is quite 
reasonable, in view of estimates based on AGN variability ($\sim 0.02$, 
Starling et al. 2004) or on the results the MHD simulations 
of the MRI (magneto-rotational) instability now accepted as the physical 
source of viscosity (again, $\sim 0.02$, Winters et al. 2003). However, the 
apparent superiority of the ADAF solution may be caused by selection effects.
The avoidance region of broad line objects is characterized by small black
hole masses and small Eddington ratio so these objects are actually very
faint.

A consistent picture of an AGN activity emerges from recent papers,
including our results. The key parameter governing the character of activity
is indeed the Eddington ratio of an object.  Specifically, 
the accretion flow in sources radiating above a per cent,
or a few per cent, of the Eddington ratio proceeds predominantly in the 
form of a cold disk even quite close to the black hole. The continuum spectra
of these sources show a prominent optical-UV  Big Blue Bump, extending in some
sources well into soft X-rays. The optical/UV emission lines are broad, and the
iron $K_{\alpha}$ line is broadened (although not necessarily extremely broad).
A natural conclusion is that all those features are related to the
existence of the cold disk down to a few hundred - a few tens of the 
Schwarzschild radii, or less. 
  
Accretion flow in sources radiating at the Eddington ratio below one per cent,
or a few per cent of the Eddington ratio do not develop features such
as a Big Blue Bump and broad emission lines (Tran 2001, 2003). 
At the same time they are 
unlikely to contain radiatively 
efficient inner flow (e.g. Merloni, Heinz and Sunyaev 2003), and 
their emission may originate predominantly as a synchrotron emission of the
jet (Merloni, Heinz \& Di Matteo 2003). 

The properties are not just bi-modal but there is also clear effect of the
Eddington ratio on all source properties even within its class, and perhaps
a smooth transition between them. Many studies
show a tight correlation of source properties with a single parameter 
(Boroson \& Green eigenvector 1, Boroson \& Green 1992; see also 
Marziani et al. (2003) and the references therein) most probably representing 
the Eddington ratio. 

As for the conclusion that the classical ADAF solution better
represents the broad line objects, we treat this result as indicative rather 
than a proof of the failure of the other models
since there are several issues which might have affected the
analysis.

\subsection{Accuracy of the theoretical relations}

\subsubsection{The size of the BLR}

In the present paper we adopted the scaling from Kaspi et al. (2000), with
$R_{BLR} \propto \lambda F_{\lambda}(5100 \AA)^{0.7}$. 
Extensive 20-year studies of NGC 5548
(Peterson et al. 2002) show that the situation is more complex due to the
dependence of the shape of the spectrum on its luminosity in this object. 
Therefore, two good representations of the source behaviour were found: 
$R_{BLR} \propto \lambda F_{\lambda}(5100 \AA)^{0.95}$ and 
$R_{BLR} \propto \lambda F_{\lambda}(1350 \AA)^{0.53}$. The 
Kaspi et al. (2000) 
formula has an index intermediate  between an optical and a UV law
since perhaps statistically most objects show weaker dependence of the 
spectral shape on luminosity than NGC 5548. Since physically we expect a
dependence of the BLR size on the ionizing flux (and not total bolometric 
luminosity) it is likely more appropriate 
to use UV flux, whenever possible, and
perhaps slightly a smaller index in size-luminosity scaling, if
statistically sound scaling is available.

\subsubsection{Evaporation radius and the accretion flow efficiency}

The version of solutions adopted in our considerations are the most
basic and simple parametric dependences, either derived from analytical
solutions or from analytical crude approximations to the numerical results.
A more careful approach does not seem necessary 
at this stage since the physical 
assumptions behind the discussed models can be questioned. Therefore,
the best support for the model came from as many observational  
tests as possible.

Liu \& Meyer-Hofmeister (2001) compared the predictions of their model with the
constraints for an inner disk radius from spectra analysis and they found
a satisfactory agreement for most objects, but Low Luminosity AGN as for 
instance NGC 4579 (see also Sect.~\ref{sect:ngc4579}) 
and M81 posed a problem for this
paper. However, the effect of magnetic field by Meyer \& Meyer-Hofmeister 
(2002) allowed for its solution.  

\subsection{Accuracy of the observational results}

\subsubsection{Intermediate Line Region}

From careful studies of the UV quasar spectra, Brotherton et al. 
(1994b) suggested that the traditional BLR consists of two components: 
one of width $\sim 2000 $ km/s and another 
very broad component of 
width $\sim 7000 $ km/s. These components also differ with respect to the net
shift in comparison to the systemic redshift. Brotherton et al. (1994b)
suggested, that the first component forms an Intermediate Line Region (ILR)
located further out and possessing a lower medium density and covering factor
than the second component, forming a Very Broad Line Region.
Similar ideas were suggested earlier by Baldwin et al. (1988) and 
Francis et al. (1992).  As is apparent from 
the profile of the MgII line in the 
composite spectra obtained from SDSS 
(Fig. 5 of Richards et al. 2002), this line might well consist of 
a narrower core and broader shoulders.

Following those ideas, we assume that the Very Broad Line Region is related
to the cold accretion disk, as the component originating closer in and
originating in material moving at a faster speed. In the analysis of MgII lines
by Forster et al. (2001) and by Kuraszkiewicz et al. (2002, 2004) both
components were identified, and we used the FWHM of the VBLR component in
our analysis. Such differentiation was not always done for other sources
used in our paper which resulted in underestimation of some line widths. 

\subsubsection{MgII vs. $H_{\beta}$ lines}

Both MgII and $H_{\beta}$ lines are Low Ionization Lines (LIL) and therefore
are supposed to arise from the vicinity of accretion disks (Collin-Souffrin
et al. 1988). However, it 
does not need to mean automatically that they form at the same distance 
from the central region. For instance, the time delay measured for MgII
lines in NGC 5548 (Clavel et al. 1992) is larger by a factor of three 
than the time delay of $H_{\beta}$ in the same source (Peterson et al. 1999). 
The scaling law of Kaspi et al. (2000) 
(see Eq.~\ref{eq:Kaspi}) is based on $H_{\beta}$ and the use of MgII lines
in principle requires an adjustment of the approach. However, McLure \& Jarvis (2002)
find the BLR scaling $R_{BLR}\propto \lambda L_{\lambda}$, if MgII
lines are used
and the continuum is measured at 3000 \AA but otherwise they do not find any
systematic difference in FWHM of MgII and $H_{\beta}$, as expected if both 
lines
are emitted at the same distance from the central ionizing source. 

Among the objects included in Table~\ref{tab1} there are two sources
with independent measurements of MgII and $H_{\beta}$ by various authors.
The measured MgII line is much broader than 
$H_{\beta}$ (PG12111+143:  5400 km/s vs. 1832 km/s; Mkn 335: 5300 km/s vs. 
1620 km/s). However, McLure \& Jarvis give 2050 km/s for the MgII FWHM of 
PG1211+143. This difference illustrates the possible problems. 
On the one hand, the data of
McLure \& Jarvis (2002) taken with IUE is of lower quality than the 
FOS/HST data analyzed by Kuraszkiewicz et al. (2002, 2004) and 
no differentiation between the ILR and VBLR components could have been
done by those authors. The Kuraszkiewicz et al. data were obtained 
with the HST, with
good S/N. On the other hand, the
modeling of MgII in NLS1 galaxies is possibly a subject of large systematic 
errors due to the presence of strong FeII emission. A specific template has to
be used in order to account for FeII before MgII modeling can be done 
(I Zw1 has been used by Kuraszkiewicz et al., following the analysis of 
Vestergaard \& Wilkes 2001) and
therefore the measurement of the FWHM of MgII may depend on this choice.  

\subsubsection{Estimate of Eddington ratio}

The estimate of the Eddington ratio in a 
given object relies both on estimates of
the source bolometric luminosity as well as the black hole mass. The method 
used in our paper is the simplest but relies on a single measured parameter
($\lambda F_{\lambda}$ at some wavelength). Still, errors do not appear large 
if we 
compare for example the mass of the black hole determined in this way with
masses from BLR monitoring. For example, logarithms of 
black hole masses for
NGC 3227 and NGC 3516 carefully determined 
by Onken et al. (2003), equal to 7.56 and 7.23 respectively, are similar to 
the values used in our paper (see Table~\ref{tab1}).
However, in principle the use of a direct method of a black hole mass 
estimation would be more appropriate since the mass and the accretion are 
independent parameters, as stressed in the study of Woo \& Urry
(2002). 

\subsubsection{Low Luminosity AGN}
\label{sect:ngc4579}

Low Luminosity AGN do not seem to pose a problem for the picture. Our sample
contains a few such objects. The well studied case of NGC 4579 is a 
good example.
Classified as a LINER, this object shows a broad component in its 
$H_{\alpha}$ line (Barth et al.
2001) with FWZI of ~18,000 km s$^{-1}$, and the FWHM can be roughly estimated
as $\sim 8000$ km/s (see their Fig.~1). We have this object in our sample (see
Table~\ref{tab1}), with the 
FWHM of the MgII line of 6800 km/s. The mass estimate
($10^6 M_{\odot}$ in our paper; $4 \times 10^6 M_{\odot}$ suggested by
Barth et al. 1996) locates this 
object relatively far to the left in our diagrams (black filled triangle),
but still within the allowed region for broad line objects. However, if the 
mass is actually much larger (Barth et al. 2001 suggest a mass an order of 
magnitude higher based on black hole mass - bulge velocity 
dispersion relation)
the object will move outside the allowed parameter space as defined by the ADAF
model. However, it will remain consistent with a disk evaporation model.
Iron line analysis in this source indicates, however, that the standard disk 
is present in this source and the source radiates at the 
Eddington ratio of $2 \times 10^{-3}$ (Terashima et al. 2000), only
by a factor of 2 lower than obtained in our analysis.
Therefore, studies of more low luminosity objects with good mass determination
would be particularly valuable.

\subsection{Other samples/objects}

\subsubsection{Low luminosity objects from the SDSS survey}

In the low luminosity broad/double peak $H_{\alpha}$ sample from SDSS 
analyzed by Strateva et al. (2003), the tail of the distribution extends up to
$\sim 16 000$ km/s. These objects cannot be included in our study and 
analyzed in the same manner as 
the other objects since these AGN usually do not show a Big Blue Bump so
Eq.~\ref{eq:masa} does not apply. However, the measured width of the lines
can be compared with constraints presented in Fig.~\ref{fig:FWHM_limits}. 
The maximum value of the FWHM of 16 000 km/s indicates that,
according to the standard ADAF and a black hole mass $M > 10^8 
M_{\odot}$ (Model A, see Fig.~\ref{fig:FWHM_limits}) 
the corresponding value of the viscosity parameter $\alpha$ is about 0.02.
If we adopt model B with $\alpha = 0.3 $ we
obtain a strong limit for the strength of the magnetic field: the parameter
$\beta$ (gas pressure to total pressure ratio) must be larger than 0.7, i.e.
the plasma must be gas pressure dominated.
The constraints from model C give the minimum value of the 
viscosity $\alpha$ as 
a function of $\beta$: $\alpha $ about 0.1 for gas pressure dominated plasma
and that could be lower if the magnetic field is important. 

\subsubsection{Objects considered by Nicastro et al. (2003)}

The upper limits for the Eddington ratio in Broad line 
objects, 0.001, obtained by Nicastro et al. (2003) locate them in 
Figs.~\ref{fig:data1} and \ref{fig:data2} possibly somewhere between 
the limits obtained from ADAF and from  disk evaporation. However, the
measurements of the FWHM of the hidden BLR was not provided by Tran (2001)
who conducted the spectro-polarimetric measurements for the sample discussed
by Nicastro et al. (2003). 

\section{Conclusions}

Tran (2001, 2003) and Laor (2003b) argue that in a 
fraction of AGN, 
broad lines are not detected - but not because they are hidden from our 
line of sight, but
because those objects do not develop a BLR.  Recent X-ray observation
of NGC 6251 supports this view (Gliozzi et al. 2004).

Our study shows that the existence of this class of objects is
a natural consequence of the evaporation of the cold disk
close to a black hole and the transition of the accretion flow into
a hot optically thin phase, if the BLR formation is intimately connected
with the cold disk. The transitions happens far away for
low Eddington ratio objects, at distances where the ionization
conditions would be correct for broad line formation, 
and the BLR apparently lose the needed disk 'support'. 

Quantitatively, the allowed parameter space (in log $L/L_{Edd}$ 
vs. log $FWHM$
and  log $M$ vs. log $FWHM$; see Figs.~\ref{fig:data1} and \ref{fig:data2})  
for broad line objects in our sample seems to be better represented by the
{\it strong ADAF principle} than by the more physical disk evaporation model.  

\begin{acknowledgements}

We are grateful to Micha\l ~~Jaroszy\' nski, Jean-Pierre Lasota and Zbyszek
Loska for very helpful discussions, and to Greg Madejski for comments to
the manuscript. This work was 
supported in part by grant 2P03D~003~22 
of the Polish State 
Committee for Scientific Research (KBN).

\end{acknowledgements}

\appendix
\section{Disk-corona mass exchange in the corona 
model with magnetic pressure component}

Disk evaporation due to conductive heating of the disk interior by the hot
accreting corona was proposed by Meyer \& Meyer-Hofmeister (1994) 
in the context
of the cataclysmic variables. This approach was further generalized in a
number od papers (Meyer-Hofmeister \& Meyer 1999, Turolla \& Dullemond 2000,
\Agata \& Czerny 2000a). The evaporation rate is calculated on the basis of
the computations of the conduction flux between the disk and the corona.
The effect depends both on the description of the conduction and the
description of the corona heating/cooling and a detailed approach to 
its vertical structure.
In accreting corona models discussed in this context, the standard assumption 
was that the heating is described by the viscosity parameter $\alpha$ and
the gas pressure in the corona, as in the description of the purely
hot (ADAF) flows. In the context of ADAF a generalization is frequently made
to include the magnetic pressure in addition to the disk pressure, since in
those models the magnetic field is needed to provide synchrotron photons
for the Compton cooling. In corona models, the synchrotron photons are
relatively unimportant since the soft disk photons are abundant but
nevertheless the magnetic component of the total pressure can also be present.
Such an effect was recently included in their model 
by Meyer \& Meyer-Hofmeister (2002).

Here we generalize the model developed by \Agata \& Czerny (2000b) in order
to include this additional magnetic pressure component.

We describe the local disk vertical structure as in \Agata \& Czerny (2000b), 
Sect. 2.2, but we assume now a more general formula for the pressure, $P$:
\begin{equation}
P = {k \over m_H \rho T_i} + P_{m} = {1 \over \beta}{k \over m_H \rho T_i},
\end{equation}
where $\beta$ is the gas pressure to the total (gas + magnetic) pressure ratio.
Gas pressure includes only the ion component, as before.

\begin{figure}
\epsfxsize = 80 mm \epsfbox[50 180 560 660]{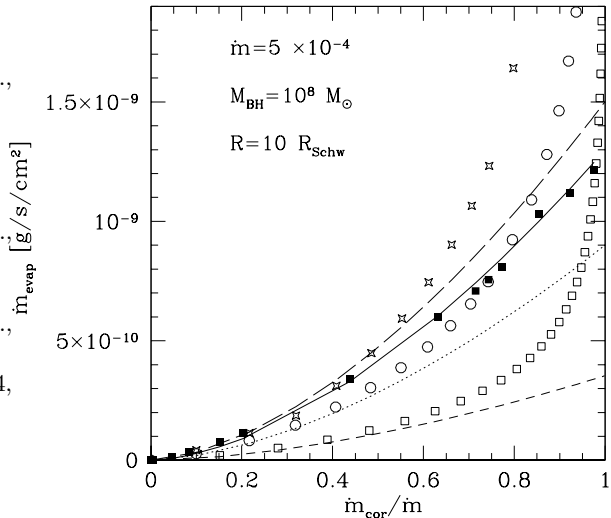}
\caption{The local relation between the accretion rate in the corona and the 
evaporation rate of the disk at $10 R_{Schw}$ for the total accretion rate 
$\dot m = 5 \times 10^{-4}$ and black hole mass $10^8 M_{\odot}$. 
Filled squares
are from  \Agata \& Czerny (2000), open circles are new 
results for $\alpha = 0.1$,
$\beta = 1$, open squares are for $\alpha = 0.1, \beta = 0.5$ and stars are for
$\alpha = 0.05, \beta = 1$.}
\label{fig:appendix1}
\end{figure}

We determine the local disk evaporation rate $\dot m_z$ as a function of the 
total (disk + corona) accretion rate, coronal accretion 
rate, the black hole mass
and the parameters $\alpha$ (viscosity) and $\beta$ (magnetic field) from
the Eq.~15 of \Agata \& Czerny (2000b). Modifying the code we found a numerical
error in the program. New results for $\beta=1$ (no magnetic pressure) and
$\alpha = 01$ are shown in Fig.~\ref{fig:appendix1}. 
The dependence on the viscosity is now much weaker than in \Agata \& Czerny 
(2000b).

The local evaporation rate shown in Fig.~\ref{fig:appendix1} increases with a 
decrease of the viscosity parameter and decreases 
with an increase of the magnetic 
field contribution to the pressure.  These trends can be understood by 
considering the relative contributions to the total pressure.  
Specifically, the total pressure at the base of the corona is 
determined by the coronal 
accretion rate under consideration and the viscosity $\alpha$. The gas pressure
there is thus proportional to $\beta /\alpha$.  
Since the ion temperature is not 
significantly lower than the virial temperature in the
coronal part of the flow, the density is mostly determined by the pressure. The
electron temperature is given by the equilibrium between the electron heating 
and cooling, and it increases with the rise in the density.
 Therefore, the electron
temperature increases with $\beta$ and decreases with $\alpha$.  
This trend determines
the trend of the integral determining $\dot m_z$ 
(equation 15 of \Agata \& Czerny 2000). 

Numerical solutions can be roughly represented by the analytical formula:

\begin{equation}
\dot m_{evap}= 1.64 \times 10^{-3} \ \alpha_{0.1}^{-1} \ \  \beta^{1.35} \ 
M_{8}^{-1} \
\dot m_{cor}^{5/3} \ R^{-3/4}
\end{equation}

This formula combined with the need for the mass conservation 
in the flow (total
accretion rate = coronal accretion rate + disk accetion rate) 
gives the solution
for the dependence of the coronal accretion rate on the radius:

\begin{equation}
\dot m_{cor}= 2.62 \times 10^2 \ \  \alpha_{0.1}^{3/2} \ \beta^{-2.025} 
 \ R^{-15/8}
\end{equation}

The condition $\dot m_{cor} = \dot m$ determines the evaporation radius
(Eq.~\ref{eq:tran} in this paper). Here we 
neglected the issue of the coronal condensation in the innermost 
region, discussed by  \Agata \& Czerny (2000).

\end{document}